\documentclass[preprint,aps]{revtex4}
\usepackage{graphicx}%
\usepackage{dcolumn}
\usepackage{amsmath}
\usepackage{longtable}
\begin{document}
\title
{ A Mixed Basis Density Functional Approach for One-Dimensional Systems
  with B-splines }
\author
{Chung-Yuan Ren$^{a,\dagger}$, 
 Yia-Chung Chang$^{b,c}$, and Chen-Shiung Hsue$^{d}$  }
\affiliation
{ $^{a}$ Department of Physics, National Kaohsiung Normal University,
Kaohsiung 824, Taiwan  	\\
$^{b}$ Research Center for Applied Sciences, Academia Sinica, Taipei 115, Taiwan \\
$^{c}$ {Department of Physics, National Cheng-Kung University, Tainan 701, Taiwan}\\
$^{d}$ Department of Physics, National Tsing Hua University,
Hsinchu 300, Taiwan \\
$\dagger$ {\it E-mail address:} cyren@nknu.edu.tw }

\begin{abstract}
A mixed basis approach based on density functional theory 
is extended to one-dimensional (1D) systems. 
The basis functions here are taken to be the localized B-splines 
for the two finite non-periodic dimensions and 
the plane waves for the third periodic direction. 
This approach will 
significantly reduce the number of the basis and therefore is computationally 
efficient for the diagonalization of the Kohn-Sham Hamiltonian.  
For 1D systems, 
B-spline polynomials are particularly useful and efficient in 
two-dimensional spatial integrations involved in the calculations
because of their absolute localization.
Moreover, 
B-splines are {\it not} associated with atomic positions when
the geometry structure is optimized, making the geometry optimization easy 
to implement.  With such a basis set we can directly
calculate the total energy of the isolated system instead 
of using the conventional supercell model with artificial vacuum regions among 
the replicas along the two non-periodic directions. The spurious Coulomb 
interaction between the charged defect and its repeated images by the 
supercell approach for charged systems can also be avoided.  
A rigorous formalism for the long-range Coulomb potential of both neutral and 
charged 1D systems under the mixed basis scheme will be derived.
To test the present method, we apply it to study the
infinite carbon-dimmer chain, graphene nanoribbon, carbon nanotube and 
positively-charged carbon-dimmer chain. 
The resulting electronic structures  are presented and discussed in details.  \\
PACS: 71.15.Mb, 73.20.-r	\\
\end{abstract}
\maketitle
\section{INTRODUCTION}
There has been increasing interest in one-dimensional (1D) 
systems on the nanoscale, such as tubes, wires, rods, ribbons, etc, because
the electronic properties of these systems are fundamentally
different from those in higher dimensions due to their unusual collective
excitations. Expectations concerning the creation and 
application of improved functional electrodevices with 
better performance characteristics are rising from 
the intensive exploration of 1D systems.
Particularly, the emergence of nanotechnology 
has led to the realization of 1D materials and 
stimulated both academic research and material innovation. 

First-principles methods based on the density functional theory have proven to
be powerful and successful in investigating the electronic structures 
and properties of solids. The use of a plane-wave basis is most natural for 
infinite 3-dimensional periodic systems, such as 
bulk solids, because of its easy implementation and the fact
that  the convergence of the calculation can be checked
systematically.  

To retain all the advantages of plane-wave expansions and of periodic boundary 
conditions in the investigation of low-dimensional systems, which are finite 
along the non-periodic direction(s),
the conventional supercell approximation is adopted by 
introducing some artificial vacuum space to separate
the periodic replica along the non-periodic direction. However,
this approach suffers from one main drawback. 
For example, in two-dimensional (2D) systems, 
it requires a vacuum layer of large thickness such that the 
interactions between the adjacent slabs are negligible, and therefore increases the number
of the plane waves along that direction. In particular, for charged systems 
(e.g. charged defects), the rather long-range tail of the Coulomb potential 
inevitably requires an extremely large separation of the two slabs and makes 
the calculation impractical \cite{LCC}.
Many correction schemes have been devised to remedy this difficulty 
\cite{TB}-\cite{RVMGR}. 
These drawbacks become even more significant in 1D systems. 

In previous work \cite{LC}-\cite{RHC}, 
a mixed planar basis approach that is conceptually simple, 
has successfully been introduced for the first-principles calculations 
of 2D systems by expanding the wavefunction along the periodic directions with
2D plane waves but, for the finite non-periodic direction, with  
1D localized basis of Gaussian functions or B-splines \cite{deBoor}.
The use of this mixed basis has several advantages over the supercell modeling:
(1) It resumes the layer-like local geometry which appears in surfaces 
and describes the wavefunction in a natural way.
(2) Because one can calculate the total
energy for an isolated slab instead of using a supercell consisting of alternating slab
and vacuum regions, the physical quantity, such as the work function can
be immediately obtained without any correction. 
(3) For charged systems, the spurious Coulomb interaction between the defect, 
its images and the compensating background charge in the supercell approach can
be automatically avoided.  (4) The number of the basis is significantly 
reduced, easing the computational burden for the 
diagonalization of the Kohn-Sham Hamiltonian.

To preserve the above good properties, 
in the present work, we extend our earlier work \cite{RHC} 
to 1D systems, i.e., with two sets of B-splines to expand the wavefunctions 
along the two finite non-periodic directions and 1D plane waves along the 
periodic one.  
B-splines are highly localized and piecewise polynomials within
prescribed break points which consist of a sequence numbers called knot
sequence \cite{deBoor} 
and have proven to be an excellent tool for the description of 
wavefunctions \cite{RHC}-\cite{RJH}. B-splines have the following properties:
(1) due to their absolute localization, the relevant matrices are sparse. This 
is particularly useful and efficient for 1D systems, 
which involve multi-dimensional spatial integrations.
(2) B-splines possess good flexibility to represent a rapidly varying
wavefunction accurately with the knots being arbitrarily chosen to
have an optimized basis. (3) B-splines are, unlike the atom-centered 
Gaussian basis, independent of
atomic positions, so the geometry optimization can be easily implemented.
Here, 
a rigorous formalism designed to treat the long-range Coulomb potential of 
both neutral and charged 1D systems within the present mixed-basis framework 
is also developed.  

We apply the present approach to study 
the infinite carbon-dimmer chain, graphene nanoribbon, carbon nanotube, and the
case of positively-charged carbon-dimmer chains. 
We perform the band structure calculation using
Vanderbilt's ultra-soft pseudopotentials (USPP) \cite{DV}.
Extensive comparisons are made to the standard supercell approach with
the popular
VASP code \cite{KJ,KF}. It is found that the calculated band structures are 
very promising but the number of the basis is significantly reduced. Aside from
the reduction, no further corrections are needed for the charged chain.  
\section{METHOD OF CALCULATION}
\subsection{B-splines}
For the sake of completeness, we
first briefly summarize the B-spline formalism.
More details can be found in Refs. \cite{RHC} and \cite{deBoor}.

B-spline of order $\kappa$ consists
of positive polynomials of degree $\kappa-1$, over $\kappa$
adjacent intervals. These
polynomials are determined by a knot sequence $\{ \tau_i\}$ and 
vanish everywhere outside the subintervals $\tau_i < s <
\tau_{i+\kappa} $.
The B-spline basis set is generated by the following relation :
\begin{equation}
B_{i,\kappa}(s)=\frac{s-\tau_i}{\tau_{i+\kappa-1}-\tau_i}B_{i,\kappa-1}(s)+\frac{\tau_{i+\kappa}-s}{\tau_{i+\kappa}
-\tau_{i+1}}B_{i+1,\kappa-1}(s),
\end{equation}
with
\begin{equation}
B_{i,1}(s)= \left \{ \begin{array}{ll}
                1,      & \tau_i \leq s < \tau_{i+1}  \\
                0,      & {\rm otherwise \ .}
                \end{array}
                \right.
\end{equation}
The first derivative of the B-spline is given by
\begin{equation}
\frac{d}{ds}B_{i,\kappa}(s)=\frac{\kappa-1}{\tau_{i+\kappa-1}-\tau_i}B_{i,\kappa-1}(s)-\frac{\kappa-1}
{\tau_{i+\kappa}-\tau_{i+1}}B_{i+1,\kappa-1}(s).
\end{equation}
Therefore, the derivative of B-splines of order $\kappa$ is simply a 
linear combination of B-splines of order $\kappa-1$, which is
also a simple polynomial and is continuous across the knot sequence.
Obviously, B-splines are flexible to accurately represent any localized 
function of $s$ with a modest number of the basis by only
increasing the density of the knot
sequence where it varies rapidly \cite{RHC}.

\subsection{Relevant matrix elements within B-spline basis}
In Vanderbilt's USPP scheme \cite{DV}, the wavefunction $\phi_i$
satisfies a secular equation of the form
\begin{equation}
H|\phi_i>=\epsilon_iS|\phi_i>
\end{equation}
under a generalized orthonormality condition
\begin{equation}
<\phi_i|S|\phi_j>=\delta_{ij}\ .
\end{equation}
Here, 
\begin{equation}
H=- \nabla^2\ + V_{pp} + V_H+ V_{xc}\ ,
\end{equation}
with $V_{pp}$, $V_H$, and $V_{xc}$ denoted as 
the pseudopotential, Hartree potential,
and exchange-correlation potential, respectively, and 
$S$ is a Hermitian overlap operator. 
In the following, we will give a detailed description of the relevant 
calculations for $H$ and $S$ operators.

By using two sets of
B-splines to describe the non-periodic $x$ and $y$ directions and 1D plane waves
for the periodic $z$ direction, the present mixed basis used to expand $\phi_i$
is defined as
\begin{equation}
< {\bf r } | { \bf \ k +  G } ; j, \kappa;j^{'}, \kappa^{'} > \
=
\frac{1}{\sqrt{L}}\
e^{i(k +  G)z }
\ B_{j,\kappa}(x)B_{j^{'},\kappa^{'}}(y)
\end{equation}
where
$ {\bf G} $ and $ {\bf k} $
denote respectively the reciprocal lattice vector, and
the Bloch wave vector.
$L$ is the length of the system along the $z$ direction.

We define
\begin{equation}
< \ B_{i,\kappa}
| \ B_{i',\kappa}>_s\ = \int ds \ B_{i,\kappa}(s)\ B_{i',\kappa}(s)
\end{equation}
where $s=x$ or $y$.
It is an integration of local polynomials with bounded support and { \it vanishes
unless the condition $|i-i'|\le \kappa$
is fulfilled }. This property is particularly useful for higher-dimensional integrals involved 
in the calculations for 1D systems.

The overlap matrix elements between two basis states are given by
\begin{equation}
< {\bf k + G} ; i, \kappa;i^{'}, \kappa^{'} \
|\ {\bf k+  G'} ; j, \kappa;j^{'}, \kappa^{'}
> \
= \
< \ B_{i,\kappa} | \ B_{j,\kappa}>_x
\ < \ B_{i',\kappa'} | \ B_{j',\kappa'}>_y
\ \delta_{ G, G'} \ .
\end{equation}

The kinetic energy matrix elements are given by
\begin{align}
&< {\bf k + G} ; i, \kappa;i^{'}, \kappa^{'} 
\ |\  - \nabla^2\  |\
 {\bf k+  G'} ; j, \kappa;j^{'}, \kappa^{'}
> \ 		\\
=& [ \
< \ {B'}_{i,\kappa} | \ {B'}_{j,\kappa}>_x \ 
< \ {B}_{i',\kappa'} | \ {B}_{j',\kappa'}>_y \  
+
< \ {B}_{i,\kappa} | \ {B}_{j,\kappa}>_x \ 
< \ {B'}_{i',\kappa'} | \ {B'}_{j',\kappa'}>_y \ 	\\
+&
< \ {B}_{i,\kappa} | \ {B}_{j,\kappa}>_x \ 
< \ {B}_{i',\kappa'} | \ {B}_{j',\kappa'}>_y \ (k+ G)^2 \
]\ \delta_{G, G'} \ .
\end{align}

$ {B'}_{i,\kappa}(s)$ is the derivative of $ B_{i,\kappa}(s)$. 
As mentioned above, 
because of the absolute localization of $ B_{i,\kappa}$ and $ B'_{i,\kappa}$, 
the evaluation for the kinetic part of $H|\phi_i>$ is only an order 
\begin{equation}
N_d(\kappa-1)\kappa \ , 
\end{equation}
where $N_d$ is the number of the basis. In practical applications, $\kappa$ 
is usually set to be 4 or 5. Therefore, the computational effort for the
construction of the kinetic energy matrix elements scales 
linearly as $N_d$.

The local part of $V_{pp}$ on each atomic site with species $\sigma$
concerned here was fitted as
\begin{equation}
V_{\text{loc}}^{\sigma} ({\bf r}) =
\ -\  \frac{Z^{\sigma}}{r}\ \text{erf}\left( \frac{r}{R_c^{\sigma}}\right)
+ \sum_i A_i^{\sigma} e^{- a_i^{\sigma}r^2}. \label{loc}
\end{equation}
Presently, the $d$-like pseudopotential is chosen as the local pseudopotential.
The first term on the right hand side of Eq. (\ref{loc})
will be referred to as the core term due to
the core charge distribution
\[
n_c( r) \ = \frac{Z^{\sigma}}{\pi^{\frac{3}{2}} {R_c^{\sigma}}^3}\
e^{- \frac{r^2}{{R_c^{\sigma}}^2}} \ .
\]

The local pseudopotential of the crystal is then given by
\begin{equation}
V_{\text{LOC}} ({\bf r}) = \sum_{{\sigma,\bf R^\sigma}}
V_{\text{loc}}^{\sigma} ({\bf r}- {\bf R^\sigma}) \ ,
\end{equation}
where ${\bf R^\sigma}$ denotes the
position of each atom with species $\sigma$.
The Fourier transform of the local pseudopotential of the crystal
excluding the core term, $ V'_{\text{LOC}} $,
is given by
\begin{align}
V'_{\text{LOC}}(\boldsymbol{\rho},g) \ 
&=\frac{1}{L}\int dz \ V'_{\text{LOC}}({\bf r})\ e^{-igz}
\\
&= \sum_{\sigma,{\bf R_\parallel^\sigma}\in\ UC}
\sum_i\left(\frac{A_i^{\sigma}}{L_u}\right) \sqrt{\frac{\pi}{a_i^{\sigma}}}
e^{-\ \frac{ g^2}{4 a_i^{\sigma}}}
\ e^{-a_i^{\sigma} (\boldsymbol{\rho-R}_\parallel^\sigma)^2}
\  .
\end{align}
Here, $ L_u$ is the length of the unit cell (UC) along $z$ axis.

The total charge distribution $n$
is defined as the sum of the core charge distributions
for all atoms in the sample $n_c$, plus the electronic charge
distributions $n_e$,
\begin{equation}
n({\bf r})= n_c({\bf r})+ n_e({\bf r}).
\end{equation}
The Coulomb potential
$V^{(C)}$ due to the total electron charge distribution and 
the exchange-correlation potential
$V_{xc}$ should be determined self-consistently.
The exchange-correlation potential are deduced from the Monte Carlo
results calculated by
Ceperley and Alder\cite{CA} and parametrized by Perdew and Zunger\cite{PZ}.
We write
\begin{equation}
V_{xc}({\bf r})
=\sum_g
V_{xc}(\boldsymbol{\rho}, g) \ e^{igz} \ .
\end{equation}

With the use of the 1D Fourier transformation of $1/r$ (see the Appendix)
\[
\frac{1}{r}
= \frac{1}{\pi} \int dq_z
K_0(|q_z| \rho)\ e^{i q_z z}\
\]
where $K_0$ is the modified cylindrical Bessel function of zero order,
the Coulomb potential due to the total charge distribution is given by
\begin{align}
V^{(C)}
= \iiint d^3 {\bf r'} \frac{n({\bf r'})}{|{\bf r}- {\bf r'}|}
\nonumber\\
=  \sum_g V^{(C)}(\boldsymbol {\rho}, g)\
e^{igz}
\label{Coul}
\end{align}
where
\begin{equation}
V^{(C)}(\boldsymbol {\rho}, g)=
2 \iint d^2 \boldsymbol {\rho'}
n({\boldsymbol {\rho'}}, g)
K_0(|g||{\boldsymbol {\rho}}- {\boldsymbol {\rho'}}|),
\label{CoulG}
\end{equation}
with 
\[
n({\boldsymbol {\rho}}, g)
= \frac{1}{L}\ \int dz \ n({\bf r}) \ e^{-i g z} \ .
\]  
$K_0$ will diverge when $g \rightarrow 0$.  
But, $V^{(C)}(\boldsymbol {\rho}, 0)$ is still well defined, as explained 
below. \\

For the $g \rightarrow 0$ case, using the asymptotic behavior of $K_0$ 
\[
K_0(x) \rightarrow -\ln x + \ln 2 - \gamma, \ \ \  x \rightarrow 0^{+}   
\]
where $\gamma$ is the Euler-Mascheroni constant, 
 $V^{(C)}(\boldsymbol {\rho}, 0)$ is now split into two terms, 
\begin{equation}
V^{(C)}(\boldsymbol {\rho}, 0)= 
2\ (\ln 2 - \gamma-\ln |g|) \iint d^2 \boldsymbol {\rho'}
n({\boldsymbol {\rho'}}, 0)
-2 \iint d^2 \boldsymbol {\rho'}
n({\boldsymbol {\rho'}}, 0)
\ln(|{\boldsymbol {\rho}}- {\boldsymbol {\rho'}}|) \ . \label{Coul0}
\end{equation}
The first term could be safely dropped if the system is charge neutral 
($\iint d^2 \boldsymbol {\rho'} n({\boldsymbol {\rho'}}, 0)=0 $). 
We demonstrate in the following that omitting such term 
still holds for charged systems.

We assume $L$, the length of the system along the $z$ direction, is 
arbitrarily large but finite. With  
\begin{equation}
n({\bf r})
=  \sum_g \ n({\boldsymbol {\rho}}, g) \ e^{ i g z} \ ,
\end{equation}
the $g=0$ component of $V^{(C)}$ will be 
\begin{equation}
V^{(C)}(\boldsymbol {\rho}, 0)=\iiint_{-L/2}^{L/2} \frac{dz'}
{ \sqrt{\Delta \boldsymbol {\rho}^2+z'^2 }} 
\ n({\boldsymbol {\rho'}}, 0) \ d^2 \boldsymbol {\rho'}  
\end{equation}
where 
$\Delta \boldsymbol {\rho}=|{\boldsymbol {\rho}}- {\boldsymbol {\rho'}}|$.
Suppose 
$L$ be much larger than the cell size in the $xy$ plane, i.e., 
$ L \gg \Delta \boldsymbol {\rho}$, then   
\begin{equation}
\int_{0}^{L/2} \frac{dz'}
{ \sqrt{\Delta \boldsymbol {\rho}^2+z'^2 }} = 
\ln \frac{L/2+\sqrt{ L^2/4+\Delta \boldsymbol {\rho}^2}}
         {\Delta \boldsymbol {\rho}} \sim 
\ln L- \ln |\Delta \boldsymbol {\rho}| \ .
\end{equation}
So, 
\begin{equation}
V^{(C)}(\boldsymbol {\rho}, 0) = 2 \ln L  
\iint d^2 \boldsymbol {\rho'} \ n({\boldsymbol {\rho'}}, 0)
-2 \iint d^2 \boldsymbol {\rho'} \ n({\boldsymbol {\rho'}}, 0) 
\ln |{\boldsymbol {\rho}}- {\boldsymbol {\rho'}}|. \label{A1} 
\end{equation}
In the case of charge neutrality, 
the first term on the right hand side of Eq. (\ref{A1}) vanishes and we 
retain Eq. (\ref{Coul0}). For charged systems with net line charge density 
$N_l =
\iint d^2 \boldsymbol {\rho'} \ n({\boldsymbol {\rho'}}, 0) \neq 0$,
such term would be huge. But clearly it is a constant that is independent upon
$\boldsymbol {\rho}$, and only causes a shift to the total energy. This kind of constant is  
irrelevant to the band structure calculation. Therefore, within the present 
mixed basis approach, we could, just like the charge-neutral case, omit the 
first term {\it without further corrections} for the charged systems.

\begin{figure}[h]
        \caption{ } \label{fig1}
\includegraphics{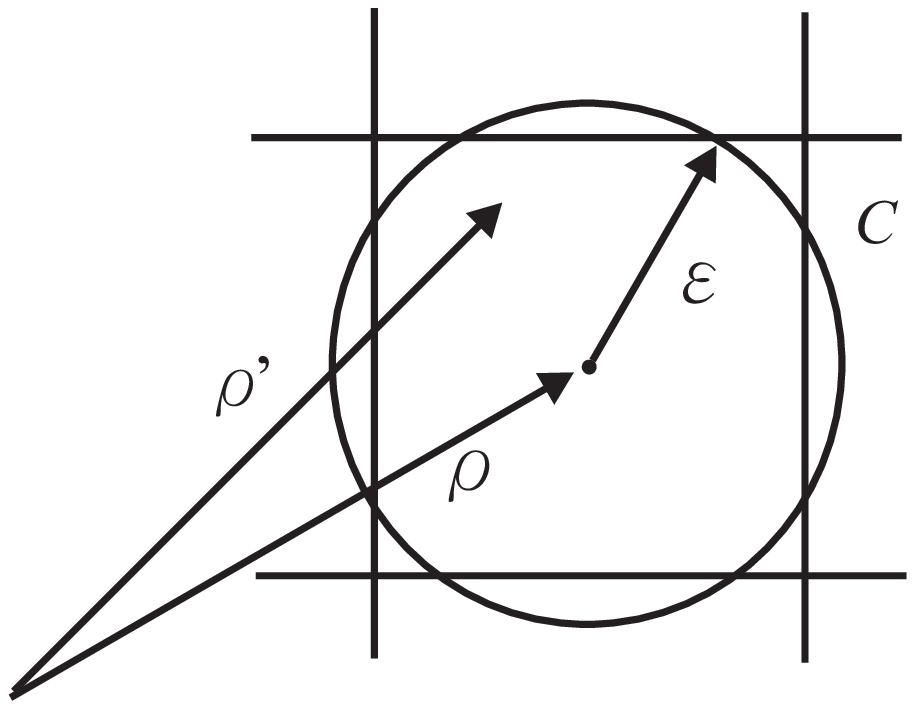}
\end{figure}

The numerical integration around the singularity at 
${\boldsymbol {\rho}}= {\boldsymbol {\rho'}}$ in Eqs. 
(\ref{CoulG}) and ( \ref{Coul0}) can be carried out and averaged over
one finer sub-grid unit. In practice, we perform such an integration over a
circle $C$, whose area is equal to that of the grid unit, as shown in Fig.
\ref{fig1}. We also assume 
$n(\boldsymbol {\rho'})$ be constant within such a circle. Then,
\begin{equation}
2 \iint_C d^2 \boldsymbol {\rho'}
n({\boldsymbol {\rho'}}, g)
K_0(|g||{\boldsymbol {\rho}}- {\boldsymbol {\rho'}}|) \
\sim \ 4\pi n(\boldsymbol {\rho},g) 
\left [ \frac{1}{g^2}-\frac{\epsilon}{|g|}K_1(\epsilon |g|) \right ],
\end{equation}
and 

\begin{equation}
-2 \iint_C d^2 \boldsymbol {\rho'}
n({\boldsymbol {\rho'}}, 0)
\ln(|{\boldsymbol {\rho}}- {\boldsymbol {\rho'}}|) 
\sim \ -n(\boldsymbol {\rho},0)\left [ 
\epsilon^2 \ln \epsilon-\frac{1}{2}\epsilon^2 \right ] .
\end{equation}
Here, $\epsilon$ is the radius of the circle $C$ and $K_1$ is
the modified Bessel function of order 1.

We define the total local potential $V_{\text{eff}}$ as
\begin{equation}
V_{\text{eff}}= V^{(C)}+ V_{xc}+ V'_{\text{LOC}}\ .
\end{equation}
To construct the $V_{\text{eff}}|\phi_i>$, we need to 
calculate 
$< {\bf k+G} ; i, \kappa;i^{'}, \kappa^{'} \ | V_{\text{eff}}|\phi_i>$
in the real-space. Again, it can be done efficiently with 
the advantage of the absolute localization of B-splines mentioned before. 

Now, lets turn to the nonlocal part of $H$. 
The atomic nonlocal potential in the Kleinman-Bylander (KB) form \cite{KB}
is given as
\begin{equation}
V_{\text{nl}}^{\sigma}({\bf r}-{\bf R^{\sigma}}) =
\sum_{nlmn'l'm'} D_{nlm,n'l'm'}^{\sigma,{\bf R^{\sigma}}}
|\beta_{n,lm}^{\sigma,{\bf R^{\sigma}}}>
<\beta_{n',l'm'}^{\sigma,{\bf R^{\sigma}}}|\ ,
\end{equation}
with
\begin{equation}
D_{nlm,n'l'm'}^{\sigma,{\bf R^{\sigma}}}=
D_{nl,n'l'}^{0,\sigma}\delta_{l,l'}\delta_{m,m'}+
\iiint d^3 {\bf r}V_{\text{eff}}({\bf r})Q_{nlm,n'l'm'}^{\sigma}
({\bf r}-{\bf R^\sigma}) \ .
\label{Dval}
\end{equation}
The projector $\beta_{n,lm}^{\sigma}$
and the augmentation function $Q_{nlm,n'l'm'}^{\sigma}$ 
vanish outside the atomic core region. 
$\beta_{n,lm}^{\sigma,{\bf R^{\sigma}}}$ denotes the projector centered at 
the ${\bf R^\sigma}$ atom, i.e.,  
$\beta_{n,lm}^{\sigma,{\bf R^{\sigma}}}=
\beta_{n,lm}^{\sigma}({\bf r} - {\bf R^\sigma})$.

The nonlocal pseudopotential of the crystal is
\begin{equation}
V_{\text{NL}} ({\bf r}) = \sum_{\sigma, {\bf R^\sigma}}
V_{\text{nl}}^{\sigma}  ({\bf r} - {\bf R^\sigma}) \ . \label{nlcry}
\end{equation}

In practical calculations, we define a box, which is large enough to
contain the core region \cite{LPCLV}.
The $\beta_{n,lm}^{\sigma}(\boldsymbol{\rho-R}_\parallel^\sigma,z)$ 
inside the box is transferred to ${\bf G}$ space, then
\begin{equation}
\beta_{n,lm}^{\sigma,{\bf R^{\sigma}}}=
\beta_{n,lm}^{\sigma}({\bf r} - {\bf R^\sigma})=
\sum_G \beta_{n,lm}^{\sigma}(\boldsymbol{\rho-R}_\parallel^\sigma,G)
\ e^{-i G R_z^\sigma}
\ e^{i G z }\ .
\end{equation}
Therefore, 
\begin{multline}
< {\bf k + G} ; i, \kappa;i^{'}, \kappa^{'} \
|\beta_{n,lm}^{\sigma,{\bf R^{\sigma}}} >  \\
= \ e^{-i(k+G)R_z^\sigma}
\sum_{G^{'}} \iint d^2 \boldsymbol{\rho} \ B_{i,\kappa}(x) B_{i',\kappa'}(y)\
\beta_{n,lm}^{\sigma}(\boldsymbol{\rho-R}_\parallel^\sigma,G')
\int dz \
\ e^{i( G'- G-k)z}\ . \label{proj}
\end{multline}
Similarly, both $ Q_{nlm,n'l'm'}^{\sigma}
({\bf r}-{\bf R^\sigma})$ and
$V_{\text{eff}}({\bf r})$ in Eq. (\ref{Dval}) are also transferred by fast 
Fourier transfer (FFT),
\begin{equation}
Q_{nlm,n'l'm'}^{\sigma}({\bf r}-{\bf R^\sigma})=
\sum_{G_h} Q_{nlm,n'l'm'}^{\sigma}
(\boldsymbol{\rho-R}_\parallel^\sigma,G_h;R_z^\sigma)
\ e^{iG_hz}\ , \label{fft1}
\end{equation}

\begin{equation}
V_{\text{eff}}({\bf r})=
\sum_{G_h} V_{\text{eff}}(\boldsymbol{\rho},G_h)
\ e^{i G_h z} \ . \label{fft2}
\end{equation}
Then, we obtain
\begin{equation}
D_{nlm,n'l'm'}^{\sigma,{\bf R^\sigma}}=
D_{nl,n'l'}^{0,\sigma}\delta_{l,l'}\delta_{m,m'}+
L_b^{\sigma} \sum_{G_h} \iint d^2 \boldsymbol{\rho} 
\left [  V_{\text{eff}}(\boldsymbol{\rho},G_h)\right ]^*
Q_{nlm,n'l'm'}^{\sigma}
(\boldsymbol{\rho-R}_\parallel^\sigma,G_h;R_z^\sigma) \ . \label{eval}
\end{equation}
$L_b^{\sigma}$ is the length of the core region box along the $z$ axis.
Note that the FFT grid density for $G_h$
in the summation of Eqs. (\ref{fft1})
and (\ref{fft2}) is not necessarily the same with
that for the wavefunction \cite{LPCLV}.
$D_{nlm,n'l'm'}^{\sigma,{\bf R^\sigma}}$ in the above should
be calculated self-consistently.
From Eqs. (\ref{proj}) and (\ref{eval}), we can evaluate  
$V_{\text{NL}}|\phi_i>$. In doing this, we need to calculate
\begin{equation}
< {\bf k+G} ; j, \kappa;j^{'}, \kappa^{'} \ |V_{\text{NL}}|\phi_i>=
\sum_{\sigma,{\bf R^\sigma}}
\sum_{nlmn'l'm'} D_{nlm,n'l'm'}^{\sigma,{\bf R^{\sigma}}}
< {\bf k+G} ; j, \kappa;j^{'}, \kappa^{'} \ 
|\beta_{n,lm}^{\sigma,{\bf R^{\sigma}}}>
<\beta_{n',l'm'}^{\sigma,{\bf R^{\sigma}}}|\phi_i>\ . 	\label{nleq}
\end{equation}
Because of the KB separation form in Eq. (\ref{nleq}), 
the computation effort for this part is also linear to $N_d$.

As for the Hermitian overlap operator $S$, which is peculiar to the Vanderbilt 
USPP scheme, it is given by
\begin{equation}
S=I+ \sum_{\sigma,{\bf R^\sigma}}\sum_{nlmn'l'm'}
q_{nlm,n'l'm'}^{\sigma}
|\beta_{n,lm}^{\sigma,{\bf R^{\sigma}}}>
<\beta_{n',l'm'}^{\sigma,{\bf R^{\sigma}}}|\ ,
\end{equation}
where $q_{nlm,n'l'm'}^{\sigma}=
\iiint d^3{\bf r} Q_{nlm,n'l'm'}^{\sigma}({\bf r})$.
$S|\phi_i>$ will be obtained similarly.
Finally, the charge density from the wave function is augmented inside the
core region,
\begin{equation}
\rho_e({\bf r}) =\sum_i [\ |\phi_i({\bf r})|^2 +
\sum_{\sigma,{\bf R^\sigma}}\sum_{nlmn'l'm'}
Q_{nlm,n'l'm'}^{\sigma}({\bf r}-{\bf R^\sigma})
<\phi_i|\beta_{n,lm}^{\sigma,{\bf R^{\sigma}}}>
<\beta_{n',l'm'}^{\sigma,{\bf R^{\sigma}}}|\phi_i>\ ].
\end{equation}

To find the lowest eigenvectors of the Hamiltonian
matrix $H$, we used the Lanczos-Krylov method developed previously 
\cite{RHC}, with  
the orthonormality of the wavefunctions maintained throughout
by the standard Gram-Schmidt orthogonalization procedure.
\section{APPLICATIONS OF PRESENT METHOD} \label{cm}
\subsection{infinite carbon-dimmer chain}
\begin{figure}[h]
        \caption{ } \label{fig2}
\includegraphics{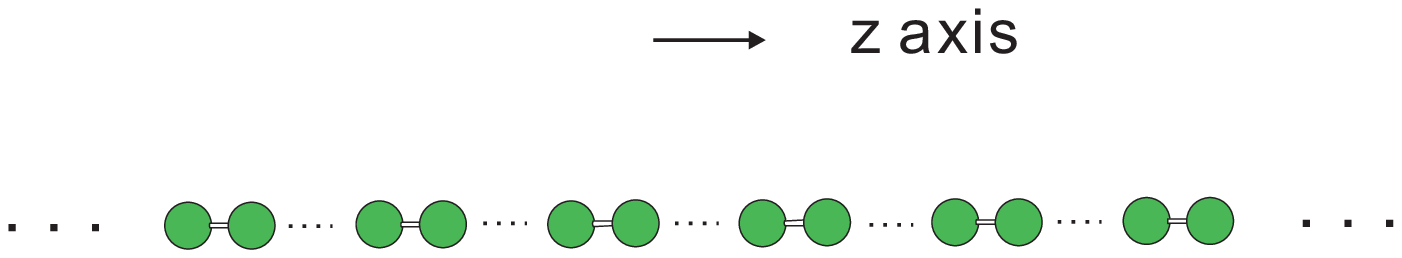}
\includegraphics{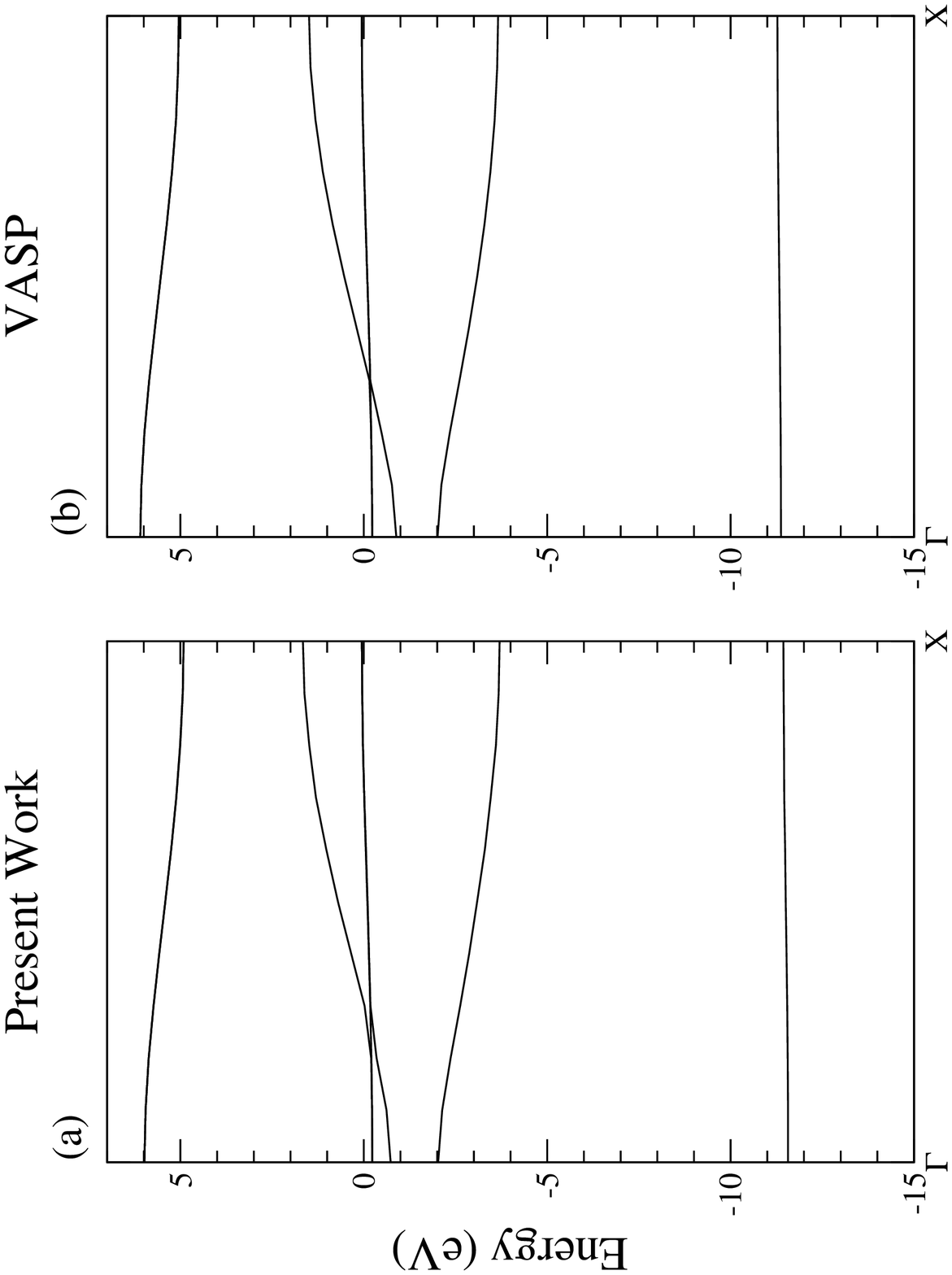}
\end{figure}

For the first example, we study a simple testing case of 
infinite carbon-dimmer chains,
as shown in the top panel of Fig. \ref{fig2}. The chain has a periodicity of 
$\sqrt 3 a_0$ ($a_0=2.461~$\AA) along the $z$ axis and the
atom distance of C-C is chosen to  $a_0/\sqrt 3$.
Two sets of 11 B-splines that each is defined over a range of 3.25 $a_0$, 
are used to expand the $x$ and $y$-component wavefunction, respectively.
The energy cutoff of the 1D plane waves along the $z$ axis is 20 Ry.
Special $k$-points of 1/8 and 3/8 (in unit of $2\pi/\sqrt 3a_0$)
were taken to sample the 1D
Brillouin zone. The C USPP was generated from
the Vanderbilt's code \cite{vancode} and its quality was   
examined previously \cite{RHC}. For comparison, we also performed the 
calculation by using the VASP code with the projector-augmented-wave potential 
(PAW) \cite{KJ,KF}.  
A typical vacuum space of 10 \AA~$\times \ 10$ \AA~ required in VASP was 
used in the calculation.
The potential is determined self-consistently until its
change is less than $10^{-6}$ Ry. 

Figure \ref{fig2}(a) displays the band structure between 
$\Gamma$ (0) and $X$ (1/2). 
The VASP counterpart is shown in Fig. \ref{fig2}(b). 
Clearly, the present calculation agrees nicely with that by VASP. 
The splitting of the twofold degenerate bands due to the symmetry of the system
is found to be smaller than 0.0001 eV. Note that the 
number of the basis is significantly reduced from $\sim 4300$ by VASP 
to $\sim 1900$ by the present method. 

%
%
\subsection{graphene nanoribbon}
\begin{figure}[h]
        \caption{ } \label{fig3}
\includegraphics{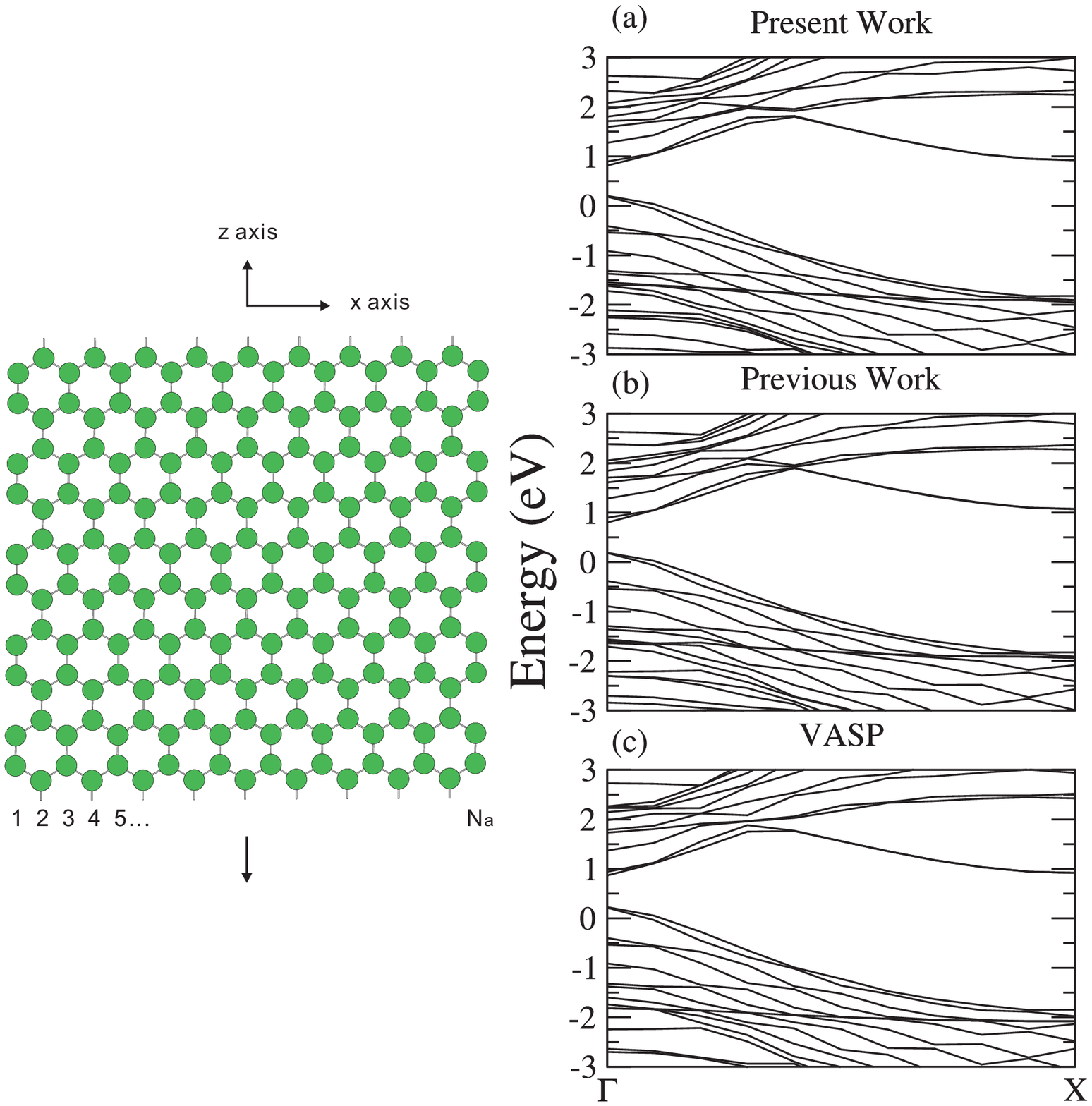}
\end{figure}

Next, in order to provide a stringent test of the present method, 
a realistic system of the armchair graphene nanoribbon is considered. 
The width of the ribbon here is chosen to include $N_a=19$ carbon dimmer lines,
as indicated in the left part of Fig. \ref{fig3}.   
One set of 73 B-splines distributed over a range of 14.25 $a_0$ and another 
of 11 B-splines over a range of 3.25 $a_0$ were used to expand 
the $x$- and $y$-components of the wavefunction, respectively. 
 The energy cutoff of the 1D plane waves for the periodic direction 
was 20 Ry. The atomic positions of the system were taken from 
Ref. \cite{RHC}. The special $k$-points for sampling the 1D Brillouin zone 
are 1/8 and 3/8. 

The present band structures near the
Fermi level are displayed in Fig. \ref{fig3}(a). To make a comparison, we also
show the previous calculations and VASP results \cite{RHC} 
in Figs. \ref{fig4}(b) and (c), respectively. In previous work \cite{RHC},  
we adopted the planar mixed basis set to study the nanoribbon 
by using the surface supercell modeling for the $x-z$ plane. Namely, we used 
only one set of B-splines to expand the $y$ component of the wavefunction. 
As compared with the VASP calculation, which was obtain by the traditional 
3-dimensional supercell modeling, the total number of the basis in previous 
work \cite{RHC} is reduced from
$\sim$17000 to $\sim$12300. Now, if we use another set of B-splines for the 
$x$ non-periodic direction, then the number is further reduced to 8800 only. 
However, it can be seen from the figures that
there is a very nice agreement between these three approaches. 
This reflects the advantage of using B-splines for non-periodic directions 
over the plane waves, especially for 1D systems. 
Actually, we have done all the calculations of the    
$N_a<19$ families. All the results are found in excellent agreements with 
those by VASP. Therefore, we
are convinced that the present program has been implemented
successfully for 1D systems and the results obtained are
very reliable. Moreover, as compared to the traditional supercell 
modeling, the use of the present mixed basis will significantly 
reduce the number of
the basis functions and speed up the calculations for 1D systems.   
\subsection{zigzag carbon nanotube}
\begin{figure}[h]
        \caption{ } \label{fig4}
\includegraphics{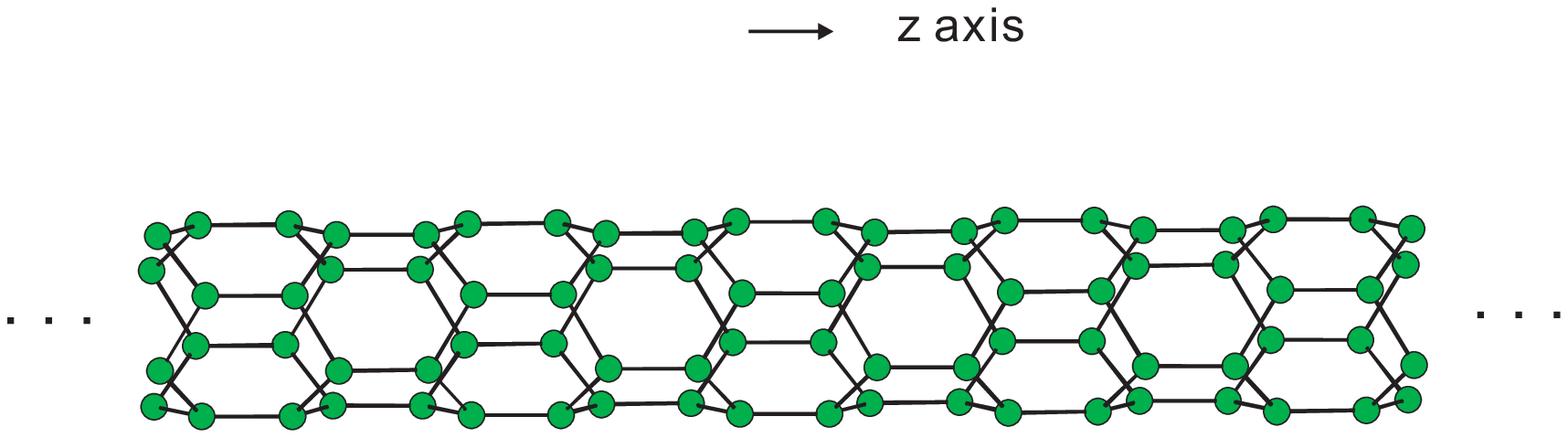}
\includegraphics{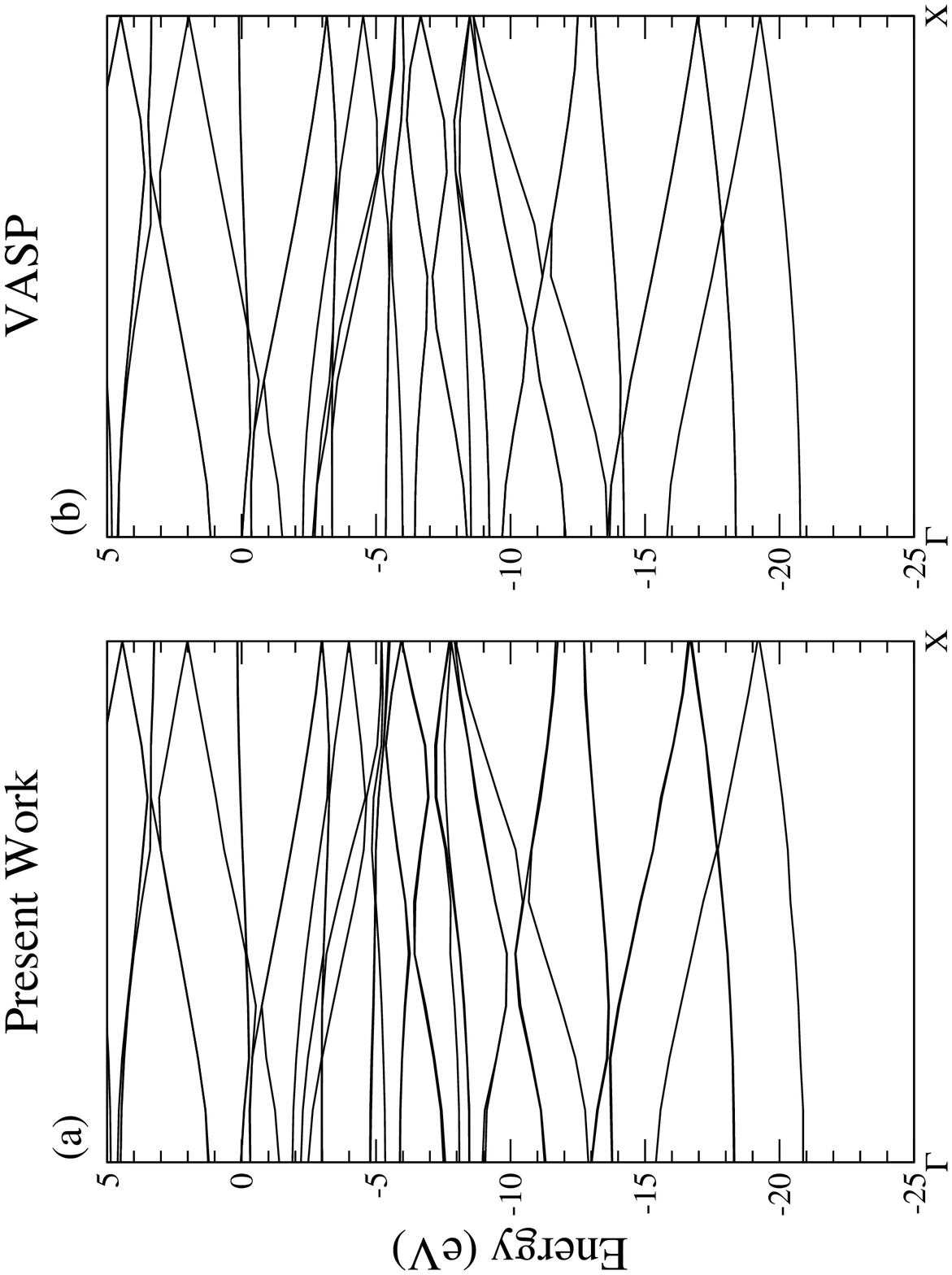}
\end{figure}

Now, we study carbon nanotubes (CNTs), allotropes of 
carbon with a cylindrical nanostructure. 
We choose the (4,0) zigzag CNT (Fig. \ref{fig4}) as the third example.  

A mixed basis set with two identical sets of 29
B-splines over a range of 6.25 $a_0$ along the two non-periodic directions 
and the plane wave cutoff of 20 Ry along the periodic direction are used.
All C atoms were kept at the ideal positions which were obtained 
by rolling the ideal graphene sheet with the C-C bond length taken to be 
$a_0/\sqrt 3$. The special $k$-point for sampling the 1D Brillouin zone 
was 1/4. The resultant band structures are presented in Fig. \ref{fig4}(a), 
along with the VASP calculations in Fig. \ref{fig4}(b) for comparison. 

Clearly, all bands compare favorably with those by VASP.
It is worth mentioning that  
it would be more efficient to investigate nanotubes or
rods if we expand the wavefunction in cylindrical coordinates 
rather than in Cartesian ones.  
Here, we utilize this system to test the present program. 
With the present algorithm to study CNT, the B-splines used 
are more densely distributed as compared to the first two cases. 
And the relevant real-space integrations were carried out with 
finer grids (48 equal divisions within $a_0$) to obtain precise results.  
Nevertheless, the total number of the basis used here is reduced from 
$\sim$9700 by VASP to $\sim$8000 by the present mixed-basis approach. 
To sum up, we have demonstrated that the present program is computationally
efficient to produced reliable results.
\subsection{charged carbon-dimmer chain}
\begin{figure}[h]
        \caption{ } \label{fig5}
\includegraphics{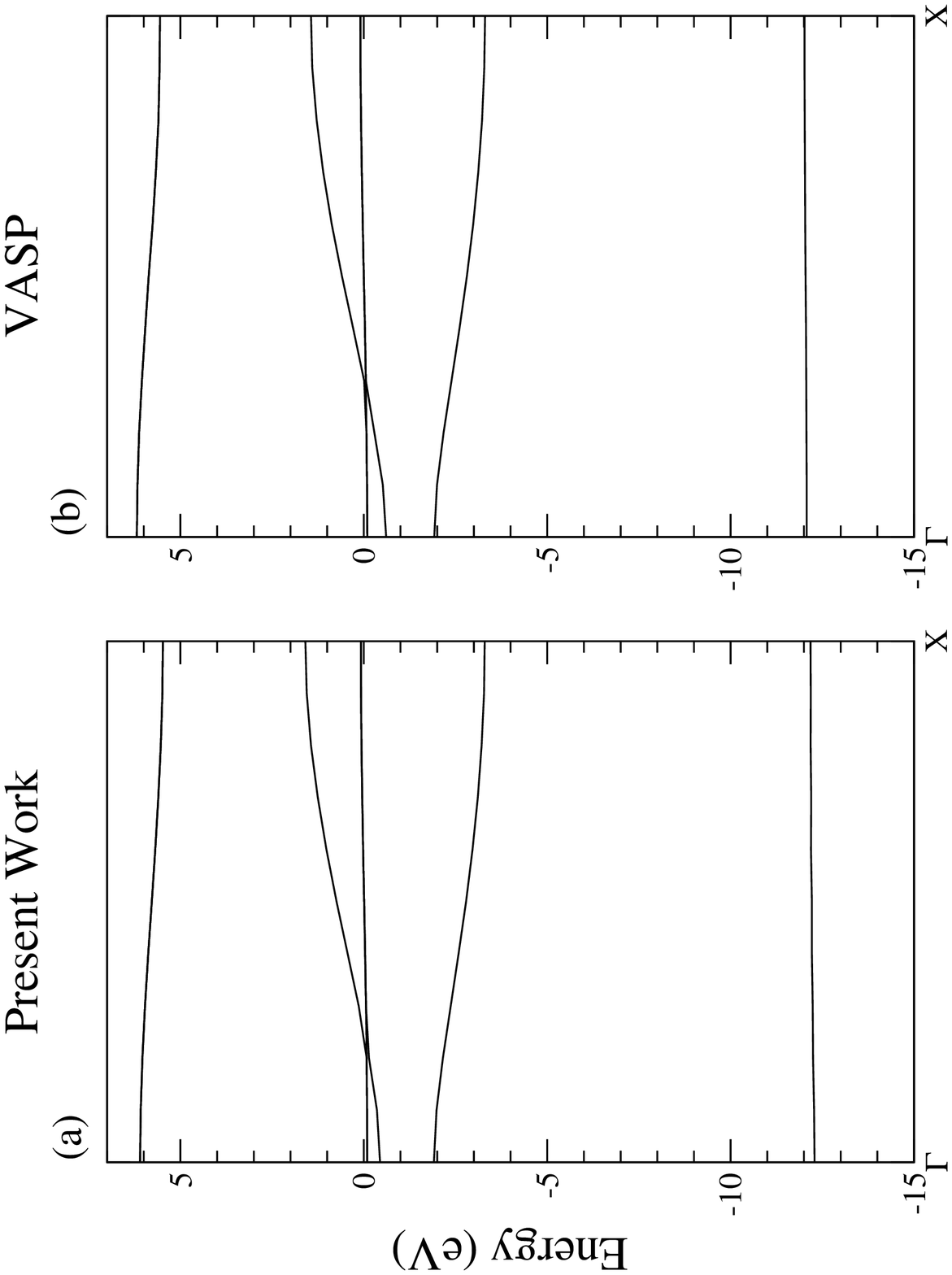}
\end{figure}

Finally, we apply the present method to charged systems which are very 
challenging for the supercell modeling because of the spurious long-range 
Coulomb interaction between the defect and its periodic images. 
For simplicity, we still use the same system of the infinite carbon-dimmer 
chain, but here with one of every eight electrons removed. 
That is, the nominal ionicity of C in this artificial positively-charged chain
is +0.5. All other computational conditions are similar to the first example 
described above. The resulting band structure is shown in Fig. \ref{fig5}(a). 
We also show the VASP result in Fig. \ref{fig5}(b), which was
obtained by using a homogeneous compensating background charge.   

It is obvious that the present approach yields very similar results to those 
by VASP. Notably, the convergence rate of the calculation 
is fast and comparable to the neutral case.   
In the supercell approach, the charged defects are unfortunately subjected to 
the spurious image interaction, and no supercell size in practice would be 
sufficient to render this long-ranged electrostatic interaction negligible.  
Various types of corrections 
have been proposed to remove the interactions
between the charged defect, its image, and the background charge 
\cite{TB}-\cite{RVMGR}. 
On the other hand, in the mixed-basis approach, it is natural to 
drop the logarithmically divergent term of  Eq. (\ref{A1})  
that has similar effect to the cancellation between the electrostatic energies
from the defects and the uniform compensating background 
assumed in the supercell model. 
No further corrections are needed in our scheme since only one single isolated
charged system rather than an array of the replicated ones is under 
consideration.   
Hence, we have developed an alternative promising method to study 
both neutral and charged 1D systems with no complications. 
\section{CONCLUSIONS} \label{con}
In conclusion, we have successfully extended the previous mixed-basis approach
to investigate the electronic structures of one-dimensional systems
with plane waves for the periodic direction and B-spline sets for
the two non-periodic directions. As compared to the existing algorithms based
upon the conventional supercell model with
alternating slab and vacuum regions, it is a real space approach along
the two non-periodic directions. Therefore, the number of the 
basis functions used to expand the wavefunction is significantly reduced and 
the spurious Coulomb interaction between the defect, its images and 
the compensating background charge appeared in the supercell approach can be 
automatically avoided.

The new technique has been demonstrated to yield accurate and computationally
efficient treatments of the infinite carbon-dimmer 
chain, graphene nanoribbon, carbon nanotube, and the system of 
the positively-charged chain.  
It is found that the band structures are all in good agreement with 
those by the popular existing codes,
but with a reduced number of basis functions. Moreover, no further corrections 
are needed for the charged case.  We have shown that the present method  
is very suitable to investigate either neutral or charged 
1D materials without the need of artificially 
large supercells or corrections for supercell interactions. 
\begin{acknowledgments}
This work was supported by Ministry of Science and Technology under 
grant numbers MOST 103-2112-M-017-001-MY3 and MOST 104-2112-M-001-009-MY2 
and by National Center for Theoretical Sciences of Taiwan.
\end{acknowledgments}

\appendix* 
\section{}
\begin{eqnarray}
\frac{1}{r}&=&\frac{4\pi}{ (2\pi)^3}\iiint \frac{e^{i{\bf q}{\bf r}}}{q^2}
 d^3 {\bf q} \\
&=&\frac{1}{ 2\pi^2}\iiint  \frac
{e^{i{\boldsymbol q_{\parallel}}{\boldsymbol {\rho}}}}{q_{\parallel}^2+q_z^2}
d^2 {\boldsymbol q_{\parallel}} \  {e^{i{ q_z}{z}}} dq_z \\
&=&\frac{1}{ 2\pi^2}\iiint  \frac
{e^{i{q_{\parallel}}{\rho}\cos\phi}}{q_{\parallel}^2+q_z^2}
q_{\parallel}dq_{\parallel}d\phi \ {e^{i{ q_z}{z}}} dq_z  \ .
\end{eqnarray}
With the identity 
\begin{equation}
J_0(x)=\frac{1}{ 2\pi} \int_0^{2\pi} e^{ix\cos\phi} d\phi 
\end{equation}
where $J_0$ is the Bessel function of order 0,
\begin{equation}
\frac{1}{r}=\frac{1}{ \pi}\iint 
\frac
{{J_0(q_{\parallel}{\rho})}}{q_{\parallel}^2+q_z^2}
q_{\parallel}dq_{\parallel} \ {e^{i{ q_z}{z}}} dq_z  \ .
\end{equation}
Using another identity 
\begin{equation}
\int_0^{\infty} \frac{xJ_0(ax)}{x^2+k^2}dx=K_0(ak)  \ \ \ [a>0, {\rm Re}\ k>0] 
\end{equation}
We obtain 
\begin{equation}
\frac{1}{r}
= \frac{1}{\pi} \int dq_z
K_0(|q_z| \rho)\ e^{i q_z z} \ .
\end{equation}
\newpage
\begin{center}
\large {\bf FIGURE CAPTIONS}    \normalsize
\end{center}
Fig. 1: Schematic plot of the grid unit containing 
        the singularity at ${\boldsymbol {\rho}}= {\boldsymbol {\rho'}}$ for
	the Coulomb potential. See text for details. \\ \\
Fig. 2: (Color online) (top) 
        Atomic structure of the infinite carbon-dimmer chain.
        (a) and (b) are the corresponding band structures obtained by
        the present work and VASP. \\ \\
Fig. 3: (Color online) (left) Atomic structure of the armchair graphene
        nanoribbon. (a), (b) and (c) are the band structures 
        near Fermi level by the present work, previous
        work \cite{RHC} and VASP, respectively.   \\ \\
Fig. 4: (Color online) (top) 
        Atomic structure of the (4,0) zigzag carbon nanotube.
        (a) and (b) are the band structures obtained by
        the present work and VASP.	\\ \\
Fig. 5: (Color online)  The band structures of the positively charged 
        carbon-dimmer chain obtained by (a) the present work and (b) VASP
        with a uniform background charge.  \\ \\
\newpage

\newpage
\newpage
\newpage
\newpage
\newpage

\begin{thebibliography}{99}
\bibitem{LCC}
A. J. Lee, T.-L. Chan, and J. R. Chelikowsky, Phys. Rev. B {\bf 89}, 075419 
(2014). 
\bibitem{TB}
S. E. Taylor and F. Bruneval, Phys. Rev. B {\bf 84}, 075155 (2011), and 
references therein. 
\bibitem{MP}
G. Makov and M. C. Payne, Phys. Rev. B {\bf 51}, 4014 (1995).
\bibitem{RVMGR}
C. A. Rozzi, D. Varsano, A. Marini, E. K. U. Gross, and A. Rubio, 
Phys. Rev. B {\bf 73}, 205119 (2006).
\bibitem{LC}
G.-W. Li and Y.-C. Chang, Phys. Rev. B {\bf 48}, 12032 (1993).
\bibitem{LC1}
G.-W. Li and Y.-C. Chang, Phys. Rev. B {\bf 50}, 8675 (1994).
\bibitem{CL}
Y.-C. Chang and G.-W. Li, Comp. Phys. Comm. {\bf 95}, 158 (1996).
\bibitem{RHC}
C. Y. Ren, C. S. Hsue and Y.-C. Chang, Comp. Phys. Comm. {\bf 188}, 94 (2015).
\bibitem{deBoor}
Carl deBoor, {\it A practical Guide to Splines}, (Springer, New York, 1987).
\bibitem{JBS}
W. R. Johnson, S. A. Blundell, and J. Sapirstein, Phys. Rev. A {\bf 37},
307 (1988).
\bibitem{JH}
H. T. Jeng, and C. S. Hsue, Phys. Rev. B {\bf 62}, 9876 (2000).
\bibitem{RJH}
C. Y. Ren, H. T. Jeng, and C. S. Hsue, Phys. Rev. B {\bf 66}, 125105 (2002).
\bibitem{DV}
D. Vanderbilt, Phys. Rev. B {\bf 41}, 7982 (1990).
\bibitem{KJ}
G. Kresse and D. Joubert, Phys. Rev. B {\bf 59}, 1758 (1999).
\bibitem{KF}
G. Kresse and J. Furthm\"{u}ller, Comput. Mater. Sci.  {\bf 6}, 15 (1996).
\bibitem{CA}
D. M. Ceperley and B. J. Alder, Phys. Rev. Lett. {\bf 45}, 566 (1980).
\bibitem{PZ}
 J. P. Perdew and A. Zunger, Phys. Rev. B {\bf 23}, 5048 (1981).
\bibitem{KB}
L. Kleinman and D. M. Bylander, Phys. Rev. Lett. {\bf 48}, 1425 (1982).
\bibitem{LPCLV}
K. Laasonen, A. Pasquarello, R. Car, C. Lee, and D. Vanderbilt,
Phys. Rev. B {\bf 47}, 10142 (1993).
\bibitem{vancode}
http://www.physics.rutgers.edu/~dhv/uspp/.
\end{thebibliography}
\end{document}